# Generalized $W_3$ Strings from Free Fields

JM Figueroa-O'Farrill, CM Hull, L Palacios, and E Ramos

*Department of Physics, Queen Mary and Westfield College, Mile End Road, London E1 4NS, UK*

**Abstract:**

The conventional quantization of $w_3$ strings gives theories which are equivalent to special cases of bosonic strings. We explore whether a more general quantization can lead to new generalized $W_3$ string theories by seeking to construct quantum BRST charges directly without requiring the existence of a quantum $W_3$ algebra. We study $W_3$-like strings with a direct spacetime interpretation— that is, with matter given by explicit free field realizations. Special emphasis is placed on the attempt to construct a quantum W-string associated with the magic realizations of the classical $w_3$ algebra. We give the general conditions for the existence of $W_3$-like strings, and comment how the known results fit into our general construction. Our results are negative: we find no new consistent string theories, and in particular rule out the possibility of critical strings based on the magic realizations.

## 1. Introduction

A $W_3$ string theory can be constructed by coupling a conformal field theory with $W_3$ symmetry and central charge $c_m = 100$ to $W_3$ gravity (see [7] for recent reviews and lists of references). Unfortunately, the only known $c_m = 100$ realizations of $W_3$ are reducible and lead to theories which are special cases of bosonic string theory. The purpose of this paper is to seek generalizations of this construction that might lead to new non-trivial $W_3$-type string theories.

In trying to maintain as far as possible the spacetime interpretation of $W_3$-strings, one is led to seek realizations of $W_3$ in terms of free bosons. The first such realization was the two-boson realization of Fateev and Zamolodchikov [3], which is the analogue of the Coulomb gas realization of the Virasoro algebra. However, it is not possible to now construct realizations with an arbitrary number of bosons by simply tensoring two-boson realizations together, because of the intrinsic nonlinearity of W-algebras, so that other constructions must be sought.

It is useful to start with a classical realization in terms of $n$ free bosons $\varphi^i$, in which

$$T_{c\ell} = \tfrac{1}{2} g^{ij} J_i J_j, \qquad W_{c\ell} = \tfrac{1}{6} d^{ijk} J_i J_j J_k \qquad (1)$$

where $J_j = i\partial\varphi_j$. These will generate a classical $\mathsf{w}_3$ algebra provided the constant tensors $g^{ij}, d^{ijk}$ satisfy certain algebraic constraints [6], which have the interpretation that the $d^{ijk}$ are the coupling constants for a Jordan algebra with cubic norm [16]. Such algebras have been classified, and consist of "generic" algebras which exist for all $n$, and the four "magic" algebras which occur for $n = 5, 8, 14, 26$, which are the anticommutator algebras of $3 \times 3$ hermitian matrices over the real division algebras: the real numbers, the complex numbers, the quaternions, and the octonions (or Cayley numbers) respectively. The generic case is reducible, constructed by adding an extra generator to a Jordan algebra with quadratic norm of dimension $n-1$, which is essentially the Clifford algebra of (some real form of) $SO(n-1)$. It is straightforward to construct a classical BRST charge for the generic and magic cases [6].

One way of proceeding is to seek a quantum operator realization of the $\mathsf{W}_3$ algebra which has this classical limit. Adding higher-derivative terms gives

$$\begin{aligned} T &= \tfrac{1}{2} g^{ij} J_i J_j + \sqrt{\hbar} a^i \partial J_i \\ W &= \tfrac{1}{6} d^{ijk} J_i J_j J_k + \sqrt{\hbar} e^{ij} J_i \partial J_j + \hbar e^i \partial^2 J_i \end{aligned} \qquad (2)$$

and the conditions for these to satisfy the $\mathsf{W}_3$ algebra were given by Romans [16]. Given a quantum $\mathsf{W}_3$ algebra, Thierry-Mieg has constructed a quantum BRST operator [17]

$$Q = \oint [cT + \gamma W + \cdots] \qquad (3)$$

which is nilpotent if and only if $c_m = 100$. The BRST operator is the fundamental ingredient in defining a consistent string theory, and its cohomology defines the physical spectrum of the theory. Moreover the algebraic structures in BRST cohomology allow one to calculate the correlation functions that, once integrated over the relevant moduli space, yield the physical amplitudes (however, a complete understanding of $\mathsf{W}_3$-moduli space is lacking at present, but see [2]).

Romans showed that for the generic classical realizations of $\mathsf{w}_3$, the constants $\{a^i, e^{ij}, e^i\}$ could be chosen so that (2) satisfy the $\mathsf{W}_3$ algebra, and these free boson realisations included ones with $c_m = 100$ that could be used to construct $W_3$ strings. However, the reducibility of the Jordan algebra led to the reducibility of the $W_3$ realisation, so that the resulting theory was a special case of the bosonic string. The bosonic strings that arise in this way are ones in which the $c = 26$ conformal matter sector consists of an Ising model together with an arbitrary $c = 25\frac{1}{2}$ conformal field theory, with the spin-three symmetry acting in the Ising sector only [8]. This was confirmed by the computation of the BRST cohomology [11], which was the culmination of a long series of attempts by a number of people (see [7] for a list of references).

The magic Jordan algebras are irreducible, however, and so would lead to a non-trivial $W_3$ string in which the W-symmetry acted on all $n$ bosons if they could be quantised. Unfortunately, there is no way of choosing the constants $\{a^i, e^{ij}, e^i\}$ to give a quantization of the magic $W_3$ algebras [12,5,14]. This is particularly disappointing, as the four magic classical theories have many attractive properties, reminiscent of the four supersymmetric Yang-Mills theories and Green-Schwarz superstring theories in $d = 3, 4, 6, 10$, which are also closely connected with the division algebras.

There have been other attempts at constructing other realizations of $W_3$ [9], but the ensuing W-string spectrum is again essentially that of an ordinary bosonic string or two. More recently, a way to obtain new W-strings from affine Lie algebras has been proposed [4], but the spacetime interpretation of these "strings without strings" is far from obvious.

One possible way out of this impasse is to try to quantize the BRST charge rather than the algebra. We start from a classical $w_3$ string with its classical BRST operator $Q$ and seek a quantum modification of $Q$ that is a nilpotent operator, but no longer demand that a quantum $W_3$ algebra exists. This is more general than the previous construction, and a BRST charge would be sufficient to allow the construction of a string theory, even though there would be no $W_3$ algebra. That such a construction might be possible is suggested by the recent work [10], in which quantum BRST charges are constructed for classical w-algebras that do not have a corresponding quantum W-algebra.

We therefore set out to find nilpotent BRST charges of the form (3) associated with "$W_3$-like" constraints of the form (2) but without requiring that the currents (2) satisfy a closed algebra. This leads to a set of algebraic conditions on the constants $\{g^{ij}, a^i, d^{ijk}, e^{ij}, e^i\}$. It follows from Thierry-Mieg's construction that these conditions will be implied by the ones found by Romans for the fields (2) to realize a quantum $W_3$-algebra; and on the other hand, the existence of a BRST operator for noncritical $W_3$-strings [1] says that these conditions will be strictly weaker. It is thus natural to ask whether other solutions exist. Our results seem to indicate that they don't, at least under a large class of Ansätze.

This paper is organized as follows. In section 2 we write down the conditions for a BRST charge to exist under the reasonable assumption that $W$ be a primary field. In this case, one can exploit conformal covariance to write down the conditions in a realization-independent fashion. In the particular case of a free-boson realization, we write down the conditions on the free parameters in (2). In section 3 we analyze these equations in several simplifying Ansätze and we find that the resulting equations have as solutions only the known BRST charges for critical and noncritical $W_3$-strings. Finally in section 4 we discuss the general picture on generalized $W_3$-strings emerging from our results.

## 2. BRST charge for $W_3$-type constraints

We start by setting up some conventions. The $W_3$-strings of interest are realized in terms of $D$ free bosons $\varphi_i$, for $i = 1, 2, \ldots, D$ obeying the following OPEs (with $\hbar = 1$)

$$\varphi_i(z)\varphi_j(w) = -g_{ij}\log(z-w) + \cdots . \tag{4}$$

We will only work with their currents $J_j(z) = i\partial\varphi_j(z)$. We take as our Ansätze the form (2) for the generators, where $g^{ij}$ is the inverse of $g_{ij}$ appearing in (4). This guarantees that $T$ generates a quantum Virasoro algebra; in other words, that the W-string background is represented by a (particular) two-dimensional conformal field theory. The central charge of this conformal field theory is given by $c_m = D - 12a^2$, where $a^2 \equiv g_{ij}a^i a^j$.

Since $T$ obeys the Virasoro algebra, the general form of the OPE between $T$ and $W$ will be

$$T(z)W(w) = \cdots + \frac{3W(w)}{(z-w)^2} + \frac{\partial W(w)}{z-w} + \cdots . \tag{5}$$

Under reasonable assumptions one can actually take $W$ to be a primary field. If we assume that in the right hand side of this OPE we find no other fields but $T$ and $W$ (and the identity), then the most general form of the OPE is then

$$T(z)W(w) = \frac{\mu}{(z-w)^5} + \frac{\lambda T(w)}{(z-w)^3} + \frac{3W(w)}{(z-w)^2} + \frac{\partial W(w)}{z-w} + \cdots . \tag{6}$$

Conformal covariance requires $\mu$ and $\lambda$ to obey $\mu = \frac{1}{2}c_m\lambda$. This means that $\widetilde{W} \equiv W - \frac{1}{4}\lambda\partial T$ is a primary field. We will see shortly that this transformation in $W$ corresponds to an automorphism of the ghost algebra. We will comment later on the possibility of other fields appearing in (5).

Therefore with no extra loss in generality we can demand that $W$ be a primary field. This imposes the following relations on the free parameters [16]:

$$e^{(ij)} = 2a_k d^{ijk} \tag{7}$$

$$e^i = \tfrac{1}{3}a_j e^{ji} \tag{8}$$

$$d^{ij}{}_j = 12e^{ij}a_j + 4a_j e^{ji} \tag{9}$$

which we can use to eliminate $e^i$ and the symmetric part of $e^{ij}$. We introduce $F^{ij} = e^{[ij]} = \frac{1}{2}(e^{ij} - e^{ji})$ and use (7) to rewrite (9) in the following form:

$$d^{ij}{}_j = 16F^{ij}a_j + 4d^{ijk}a_j a_k \tag{10}$$

which explicitly involves the $\frac{1}{6}D^3 + D^2 + \frac{5}{6}D$ free parameters in our theory: $d^{ijk}$, $F^{ij}$, and $a^i$.

In order to construct the BRST charge we introduce two fermionic ghosts systems: $(b, c)$ and $(\beta, \gamma)$ of conformal weights $(2, -1)$ and $(3, -2)$ respectively, with OPEs

$$b(z)c(w) = \frac{1}{z-w} \qquad \text{and} \qquad \beta(z)\gamma(w) = \frac{1}{z-w} \ . \tag{11}$$

The BRST charge will have the generic form (3), where the $\cdots$ means terms which are at least trilinear in the ghosts. Let us first show that we can take $W$ to be primary without extra loss of generality. Substitute $W = \widetilde{W} + \frac{1}{4}\lambda\partial T$ in the expression (3), where $\widetilde{W}$ is primary. Integrating by parts, we can rewrite the linear terms in the BRST current as $\widetilde{c}T + \gamma\widetilde{W}$, where $\widetilde{c} = c - \frac{1}{4}\lambda\partial\gamma$. Let us now complete this transformation by defining $\widetilde{b} = b$, $\widetilde{\gamma} = \gamma$, and $\widetilde{\beta} = \beta - \frac{1}{4}\lambda\partial b$. This transformation can be checked to be an automorphism: that is, the OPEs of the tilded ghosts is again given by (11). The net result, however, is that in terms of the tilded variables (but dropping the tildes) the BRST charge still satisfies the Ansätze (3) where now $W$ is a primary field of conformal weight 3.

We will find it convenient to break up the BRST charge in a slightly different way than the one suggested by (3). Exploiting the fact that the algebra respects the $(b, c)$ and $(\beta, \gamma)$ ghost numbers separately, we decompose the BRST charge by $(\beta, \gamma)$ ghost number. We shall refer to this ghost number as the "degree." It is easy to see that it can only have terms of degrees 0, 1, and 2: $Q = Q_0 + Q_1 + Q_2$. Because the algebra respects the degree, the condition $Q^2 = 0$ breaks up into a set of conditions: $Q_0^2 = 0$, $[Q_0, Q_1] = 0$, $Q_1^2 + [Q_0, Q_2] = 0$, $[Q_1, Q_2] = 0$, and $Q_2^2 = 0$. The last condition is automatically satisfied since there are no operators we can write down with the correct combination of degree, conformal weight, and ghost number.

Let us first take a look at $Q_0$. It has the form

$$Q_0 = \oint [cT + \cdots] \ , \tag{12}$$

where $\cdots$ stands now for terms which are of degree 0, ghost number 1, conformal weight 1 and at least trilinear in ghosts. Because $T$ obeys the Virasoro algebra, this should correspond to the Virasoro BRST operator where we treat the $(\beta, \gamma)$ ghosts as conformal matter. Indeed, if one writes the most general form for $\cdots$ in (12) above and imposes that $Q_0^2 = 0$, one finds that $Q_0$ must have the form

$$Q_0 = \oint c\left(T + T_{\beta\gamma} + \tfrac{1}{2}T_{bc}\right), \qquad (13)$$

with the energy-momentum tensors $T_{bc}$ and $T_{\beta\gamma}$ given by

$$\begin{aligned} T_{bc} &= -2b\partial c - \partial bc \\ T_{\beta\gamma} &= -3\beta\partial\gamma - 2\partial\beta\gamma, \end{aligned} \qquad (14)$$

and where in addition the central charge of the matter realization is fixed to $c_m = 100$. In other words, this constrains the free parameters $a^i$ to obey the quadratic relation

$$a^2 = \frac{D - 100}{12}. \qquad (15)$$

The condition $[Q_0, Q_1] = 0$ restricts the form of $Q_1$ to be simply $Q_1 = \oint \gamma W$, which allows us to relate $Q_1^2$ directly to the algebra satisfied by $W$. Since $W$ is a primary field, conformal covariance dictates that it obeys the following OPE

$$\begin{aligned} W(z)W(w) &= \frac{\tfrac{100}{3}}{(z-w)^6} + \frac{2T(w) + X(w)}{(z-w)^4} + \frac{\partial T(w) + \tfrac{1}{2}\partial X(w)}{(z-w)^3} \\ &\quad + \frac{1}{(z-w)^2}\left[Y + \tfrac{1}{17}TX + \tfrac{9}{68}\partial^2 X + \tfrac{16}{261}\Lambda + \tfrac{3}{10}\partial^2 T\right](w) \\ &\quad + \frac{1}{z-w}\left[\tfrac{1}{2}\partial Y + \tfrac{1}{34}\partial(TX) + \tfrac{5}{204}\partial^3 X + \tfrac{8}{261}\partial\Lambda + \tfrac{1}{15}\partial^3 T\right](w), \end{aligned} \qquad (16)$$

where $X$ and $Y$ are arbitrary primary fields of weights 2 and 4 respectively, and where $\Lambda = (TT) - \tfrac{3}{10}\partial^2 T$. In the absence of $X$ and $Y$, the above algebra is of course nothing but $\mathsf{W}_3$ at $c_m = 100$.

Because $[Q_0, Q_1] = 0$, it also follows that $[Q_0, Q_1^2] = 0$. Only when $Q_1^2 = -[Q_0, Q_2]$ for some $Q_2$ does the BRST charge $Q$ exist. We must therefore write down the most general $Q_2$ and check this condition. It is straightforward to conclude that the dependence of $Q_2$ on the $J_i$ can only occur via the combinations present in $X$ and $Y$. Moreover it follows that in order to compute $[Q_0, Q_2]$ we only need to know the conformal properties of the fields $X$ and $Y$. A simple computation shows that $Q_1^2 = -[Q_0, Q_2]$ provided that $Y = 0$ and that $Q_2$ is given by

$$Q_2 = \oint \tfrac{1}{2}\left[\tfrac{50}{783}\gamma\partial^3\gamma b - \tfrac{25}{261}\partial\gamma\partial^2\gamma b + \gamma\partial\gamma b\left(\tfrac{16}{261}T + \tfrac{1}{17}X\right)\right]. \qquad (17)$$

With this choice of $Q_2$ it is straightforward to show that $[Q_1, Q_2] = 0$ independent of the OPE of $W$ and $X$.

In summary, let us first point out that this result is in fact independent of the realization. In other words, we can associate a BRST operator to the algebra of $T$ and $W$ whenever $T$ and $W$ satisfy the deformation (16) of the $\mathsf{W}_3$ algebra. In our particular bosonic realization, the only constraint on our parameters is that the primary field $Y$ appearing in the second-order pole of the OPE (16) vanish. This constraint can be written down in terms of the parameters entering into our realization. For notational convenience we introduce the parameters $K^{ij}$ related to the primary field $X(z)$ as follows:

$$\tfrac{1}{17}X(z) = \tfrac{1}{2}K^{ij}J_iJ_j(z) + K^{ij}a_i\partial J_j(z) , \tag{18}$$

where, in addition, the trace of $K$ is constrained by $K^i{}_i = 12K^{ij}a_ia_j$. The $K^{ij}$ are to be found simply by comparing the fourth-order pole of $W(z)W(w)$ and of (16). One finds that $K^{ij}$ is given by

$$\begin{aligned}K^{ij} = & -2g^{ij} - 12F^i{}_kF^{jk} + d^{ikl}d^j{}_{kl} - 12F^{(i}{}_kd^{j)kl}a_l \\ & +4d^{ij}{}_kF^{lk}a_l + 2d^{ij}{}_kd^{klm}a_la_m - 3d^i{}_{kl}d^{jk}{}_ma^la^m .\end{aligned} \tag{19}$$

It is then a simple matter to compute the second-order pole of $W(z)W(w)$ and set it equal to the second-order pole of (16). Doing so we find the following equations:

$$d^{(ij}{}_rd^{kl)r} = \tfrac{16}{261}g^{(ij}g^{kl)} + g^{(ij}K^{kl)} \tag{20}$$
$$d^{ij}{}_lF^{lk} - F^{l(i}d^{j)k}{}_l + \tfrac{1}{2}d^{k(i}{}_ld^{j)lm}a_m = \tfrac{8}{261}g^{ij}a^k + \tfrac{1}{4}g^{ij}K^{kl}a_l + \tfrac{1}{4}K^{ij}a^k \tag{21}$$
$$F^i{}_kF^{jk} - \tfrac{1}{4}d^{ik}{}_ld^j{}_{km}a^la^m = \tfrac{49}{522}g^{ij} + \tfrac{3}{4}K^{ij} \tag{22}$$
$$F^{(i}{}_kK^{j)k} = \tfrac{1}{2}d^{ij}{}_kK^{kl}a_l - \tfrac{1}{2}K^{(i}{}_kd^{j)kl}a_l \tag{23}$$
$$d^{ijk}K_{jk} = 16F^i{}_kK^{jk}a_j + 4d^{ij}{}_kK^{kl}a_ja_l . \tag{24}$$

As a check when we put $K^{ij} = 0$ we recover the equations obtained in [16] for $T$ and $W$ to form a quantum $\mathsf{W}_3$ algebra at $c_m = 100$.

These equations as they stand are rather complicated and their general solution has so far eluded us. We will however study them in the next section under the simplifications afforded by several reasonable Ansätze.

### 3. Some Ansätze

*3.1. Quantization of the classical BRST charge*

The realization-independent condition found in the previous section for the existence of a BRST charge of the form (3) forces the classical algebra of $T$ and $W$ to be a soft algebra; that is, that the multiplicative ideal generated by $T$ and $W$ be a Poisson subalgebra. Or in other words, that the constraints (here $T$ and $W$) be of the first class. This is a necessary condition for the classical BRST charge to exist; the results of the next section are consistent with the idea that $Q$ in (3) should be a deformation of a classical BRST charge.

Given currents (2) satisfying the OPEs (5) and (16), the corresponding classical currents (1) satisfy a classical (Poisson bracket) algebra that includes

$$\{W_{c\ell}(z), \quad W_{c\ell}(w)\} \\ = [T_{c\ell}(z)T_{c\ell}(w) + (T_{c\ell}(z)X_{c\ell}(w) + T_{c\ell}(w)X_{c\ell}(z))] \cdot \delta'(z-w) , \quad (25)$$

after suitable rescalings of the currents. In addition, $W_{c\ell}$ and $X_{c\ell}$ are classical tensors of weights 3 and 2, respectively, relative to the centerless Virasoro algebra generated by $T_{c\ell}$. This is the classical algebra $\mathsf{w}_3$ if and only if $X_{c\ell} = 0$; otherwise it is a soft algebra.

As a first Ansatz, we demand that the BRST charge be a quantum deformation of the classical BRST charge [6] of a classical $\mathsf{w}_3$ algebra of the form (3) where the corrections are analytic in $\sqrt{\hbar}$. This requires $X_{c\ell} = 0$, which is equivalent to $K^{ij} = 0$, and from our comments in the previous section, the resulting quantum BRST charge is the one found by Thierry-Mieg. In other words, if we insist in retaining $\mathsf{w}_3$ as a classical limit, the quantization of the BRST charge is equivalent to the existence of a quantum realization of $\mathsf{W}_3$. One can show that this holds true even if we don't assume that $W$ is a primary. In fact, $W$ is forced to be primary by the resulting conditions. In particular, this tells us that the classical BRST charge for the magic realizations of Romans have no nilpotent quantum deformation and hence that they play no role as building blocks for critical $\mathsf{W}_3$-like strings.

*3.2. Coupling to $\mathsf{W}$-Liouville Theory*

Next, it is natural to ask if the magic realizations can still play a role in the non-critical case [1]. That is, we can try and construct W-strings based on a matter sector which is a magic realisation of $\mathsf{w}_3$, dressed with suitable $\mathsf{W}_3$ Liouville fields. In this case, the conditions for the nilpotency of the BRST charge become slightly weaker and, in particular, the central charge of the matter sector is no longer bound to be equal to 100.

It is simple to check that the magic realizations can be used to construct a "classical" $\mathsf{W}_3$-Liouville BRST charge, obtained as follows: let us consider two commuting $\mathsf{w}_3$ algebras with generators $T_M, W_M$ and $T_L, W_L$ respectively. The linear combinations

$$T = T_M + T_L \qquad (26)$$

$$W = W_M + iW_L \tag{27}$$

obey a soft algebra of the form (25) with $X_{c\ell}$ equal to $-T_L$, and it is now straightforward to construct its associated BRST charge. The name is justified by noticing that the algebra defined by $T$ and $W$ is a contraction of the one defined in [1] used to construct the BRST charge associated with non-critical strings. Since the only relevant ingredients in the previous construction are the two commuting w₃ algebras, any of the Romans solutions (magic or generic) is as good as any other. As we will show below that is no longer true in the quantum case, which singles out the generic solutions as the only consistent ones.

From the point of view of our free field realisations, this Ansatz boils down to the consideration of a diagonal $K^{ij}$ with only two distinct eigenvalues, which can be taken to be, without loss of generality, 1 and $-1$, and $d$-symbols without mixing terms between the two sectors. It is clear from (20) that in this case the "matter" and "Liouville" bosons have to obey separately the Romans conditions. If we choose the generic solution of these equations we find a solution that corresponds to a free field realization of the OPEs in [1] corresponding to two commuting W₃ algebras. If we now consider a magic realization for, let us say, the matter sector, it is a straightforward computational matter to check that there are no solutions for any values of the $a^j$, $e^j$, and $e^{ij}$ because the equations for this sector again collapse to the conditions obtained by Romans for the quantization of the algebra.

One can also try some further Ansätze in which the above conditions on the $d$-symbols are relaxed to admit some mixing. In particular one can demand that the Liouville modes only appear in the mixed $d$-symbols through their energy-momentum tensor $T_L$; or, perhaps less justifiably, that the magic bosons only couple to the Liouville modes via their energy-momentum tensor $T_M$. In both cases, the unique solution demands that there is again no mixing.

From a more general standpoint, it is natural to ask if it is possible to generalize the noncritical solution to $K^{ij}$ with more than two distinct eigenvalues, while still keeping the condition that there is no mixing among the different sectors at the level of the $d$-symbols. We have checked that this is not the case. All of this seems to indicate the rigidity of the known solutions.

## 4. Conclusions

Although our results are not conclusive, they shed serious doubts about the possibility of constructing new W₃-string models from of free fields that are inequivalent to special cases of critical or non-critical bosonic strings. Of course, a complete treatment would require finding the solution to the system of equations derived in section 2, a challenge that so far we have been unable to meet. Nevertheless, we have shown that the natural candidates for a nontrivial $W_3$-string, which are provided by the so-called magical realizations of the classical $W_3$ algebra, do not yield a nilpotent BRST charge at the quantum level in either the critical or noncritical scenarios.

This does not imply that there are not physically interesting $W_3$-strings with a spacetime interpretation, but rather that if they do exist they do not have a simple free field description. In fact, some hints in that direction are provided by the connection of the scale invariant rigid particle and $W_3$ [15]. It then seems plausible to conjecture that some kind of rigid string[1] might yield an algebra of constraints isomorphic to $W_3$; if so, this would open a new approach to this fascinating subject.

## Acknowledgement

One of us (LP) would like to thank the ICTP, Trieste, for its hospitality while the last stages of this work were carried out.

This paper is archived as `hep-th/9409129`.

---

[1] This type of string has been proposed by Polyakov as an equivalent description of QCD in four dimensions [13].